\documentclass{mn2e}
\usepackage{psfig}
\usepackage{epsf}
\newcommand{\ltaraw}{$\; \buildrel < \over \sim \;$}
\newcommand{\lta}{\lower.5ex\hbox{\ltaraw}}
\newcommand{\gtaraw}{$\; \buildrel > \over \sim \;$}
\newcommand{\gta}{\lower.5ex\hbox{\gtaraw}}

\newcommand{\pks}{PKS 1830-211}

\loadboldmathitalic  
\title [Microlensing induced absorption line variability]
{Microlensing induced absorption line variability}
\author[G. F. Lewis \& R. A. Ibata]
{Geraint F. Lewis$^{1,2}$ \& R. A. Ibata$^{3}$\\
$^{1}$
Anglo-Australian Observatory, P.O. Box 296, Epping, NSW 1710, Australia:
Email \tt{gfl@aaoepp.aao.gov.au}\\
$^{2}$
Present Address: Institute of Astronomy,
School of Physics, University of Sydney, NSW 2006, Australia:
Email {\tt gfl@physics.usyd,edu.au}\\
$^{3}$
Observatoire de Strabourg, 11, rue de l'Universite, F-67000, Strasbourg, 
France:
Email \tt{ibata@pleiades.u-strasbg.fr}
}
\date{\today}
\begin{document} 
\maketitle 
\begin{abstract}
Gravitational microlensing has  proven to be a powerful  probe of both
the structure at the heart of quasars and the mass function of compact
objects in  foreground lenses.  This  paper examines the  potential of
gravitational  microlensing  in  probing  the scale  of  structure  in
absorbing material  within the lensing  galaxy. We find that,  in this
high optical  depth regime,  significant variations in  the equivalent
width of absorption  features can be induced, although  the details of
these  are dependent  upon the  scale  of structure  of the  absorbing
material.  The  paper concludes with an examination  of the absorption
line  variability  observed   in  the  gravitationally  lensed  quasar
PKS1830-211, demonstrating how this may indicate the presence of small
scale structure in  the cold molecular gas present  within the lensing
galaxy.
\end{abstract}

\section{Introduction}\label{introduction}
With  the  advent  of  dedicated  monitoring  programs,  gravitational
microlensing of quasars  has provided some of the  strictest limits on
the scale  of the central accretion  disk (e.g. Wyithe,  Agol \& Fluke
2002) and  extended emission  line material  (Lewis, Irwin,  Hewett \&
Foltz 1998).  As well as  imaging the source, microlensing also probes
the nature of the lensing  galaxy, with the statistics of microlensing
variability constraining the mass  function of the microlensing masses
in the intervening system (Wyithe et al. 2000; Wyithe \& Turner 2001).

In this paper we demonstrate how microlensing provides a further probe
of the distribution  of material within a lensing  galaxy. Rather than
focusing  upon  the point-like  microlensing  masses,  we examine  the
influence of material  responsible for producing absorption signatures
in the spectrum of the more distant quasar.  This absorption signature
is time dependent, as relative  movement of the microlenses results in
differing microimage brightnesses.

In Section~\ref{microsection} we outline the general approach, with an
extension  into   the  high  optical  depth   microlensing  regime  in
Section~\ref{high}.  Sections~\ref{clouds}  and \ref{magmaps} describe
the cloud  distributions employed in this study  and the magnification
maps  from which  the results  are taken.   Section~\ref{pks} examines
whether the  radio spectral variability  observed in the  lensed radio
quasar  \pks\ is  consistent  with the  hypothesis  presented in  this
paper,  while   the  conclusions  to  this  study   are  presented  in
Section~\ref{conclusions}.

\section{Microlensing through variable absorption}\label{microsection}
The  most  dramatic  influence  of  a  microlens  as  is  crosses  the
line-of-sight to  a distant source  in the resultant variation  in the
apparent brightness  of that source.   Such variability has  been seen
for not only  microlenses within our own Galaxy  (Alcock et al. 1993),
but  also in  microlensing of  distant  quasars (Irwin  et al.   1989;
Corrigan et al. 1991).

If galaxies were comprised  solely of smoothly distributed matter then
gravitationally lensed images would  be just single distorted versions
of the  source. Due to the  granularity of matter,  distributed on the
small scale in compact objects  such as stars, a closer examination of
gravitationally lensed images  would reveal them to be  comprised of a
myriad of  microimages.  These are distributed on  scales smaller than
current imaging resolution and only  a single composite macro image is
observed.  These  images possess a distribution  of brightnesses which
is dependent  upon the relative  positions of the  microlensing masses
and  the projected position  of the  source.  The  brightness weighted
mean position of the resulting light distribution will not necessarily
lie  at  the   position  of  the  macroimage  in   the  smooth  matter
lens. Hence, as microlensing is a temporal phenomenon, the movement of
a microlensing  mass produces a time varying  astrometric offset.  For
microlensing  events  in  the  Galactic  halo  (Paczynski  1998),  the
astrometric  shifts are of  the order  of several  milliarcseconds and
will   be  readily   detectable   with  the   launch  of   space-based
interferometers such as SIM ({\tt http://sim.jpl.nasa.gov/}).

\begin{figure}
\centerline{ \psfig{figure=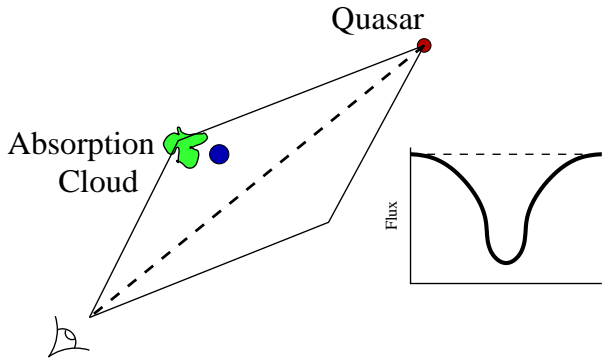,angle=270,width=3.5in}}
\caption[]{  An   illustration  of  the   influence  of  gravitational
microlensing on  absorption profiles.  The dashed  line represents the
unperturbed  path  light  would  take  from a  distant  quasar  to  an
observer. This line-of-sight is  unobscured and no absorption features
appear in  the quasar spectrum  (schematically depicted as  the dashed
spectrum in the inset).  When a microlens, the dark circle, lies close
to the line-of-sight,  light can take multiple paths  from a source to
an  observer,   and  these  paths   do  not  necessarily   follow  the
unmicrolensed  path.   In the  example  above,  two  paths around  the
microlens are  illustrated, one of  which impinges upon  an absorption
cloud. This line-of-sight now carries an absorption signature from the
cloud. While the multiple  lines-of-sight result in multiple images on
the sky,  these are  unresolved during microlensing,  and so  a single
quasar image is seen. The spectrum of this quasar image, however, also
displays  the absorption feature,  weighted by  the brightness  of the
various microimages, as represented schematically in the inset. }
\label{fig1}
\end{figure}

Typically  it is  assumed that  the various  light paths  pass through
empty space, or through a smooth matter component that does not absorb
the  light.    As  well   as  stars,  galaxies   contain  continuously
distributed material, namely gas and dust. This material is clumped on
a  range of  scales,  with indications  that  the interstellar  medium
exhibits   fractal  structure  (Elmegreen   1997).   Figure~\ref{fig1}
presents   a  (very)   schematic  example   of   microlensing  through
inhomogeneous absorption material; the dashed line depicts the path of
light from the source to an  observer in the absence of a microlensing
mass.  In this example, this  path does not impinge upon any absorbing
material  and  the  spectrum  of  the  source  does  not  possess  any
absorption feature,  as represented  by the dashed  line in  the inset
spectrum.   With  a  microlensing  mass close  to  the  line-of-sight,
however, the light  from the source follows one of  two paths from the
source to the  observer. With this example, one  of these paths passes
though  a cloud  of  absorbing material,  resulting  in an  absorption
imprint.  As the individual  microimages are unresolved, the resultant
spectrum will be the sum of the light from the two paths, one of which
possesses an absorption trough; this  is illustrated as the solid line
in the inset  spectrum. As microlensing is a  temporal phenomenon, the
spectrum  of the source  will begin  unabsorbed.  As  the microlensing
mass passes through the line-of-sight,  the absorption will be seen to
appear and change, until finally the microlensing mass moves away from
the line-of-sight and the spectrum returns to its unabsorbed state.

\begin{figure}
\centerline{ \psfig{figure=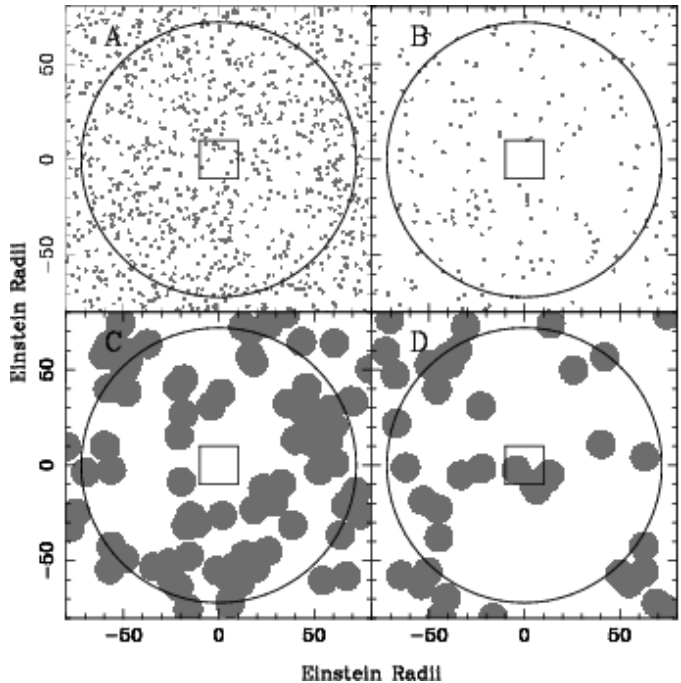,angle=270,width=3.5in}}
\caption[]{Examples of the cloud distributions employed in this study;
these specifically  represent the  clouds used for  model set  2.  The
grey denotes  the extent  of the absorption  clouds. The  large circle
represents the region over which  the microlensing stars are scattered,
whereas the small central square is  the region in the source plane in
which the traced rays are collected and the microlensing magnification
maps is calculated.  The labels  in the upper left-hand corner of each
frame refer  to the differing  cloud sizes and densities  employed, as
outlined in Section~\ref{clouds}.
\label{fig3}}
\end{figure}

Seen  at cosmological distances,  however, microlensing  by individual
stars  is unlikely  to show  such an  effect.  This  can be  seen when
considering the physical  separation of the rays of  light through the
lens. When there is a good  alignment between the source and lens, the
resulting images undergo a  similar magnification and are separated by
roughly twice the Einstein Radius, hence
\begin{equation}
Separation \sim 2. \sqrt{ 4\frac{ G M }{c^2}\frac{D_{ol}D_{ls}}{D_{os}}} 
\label{einstein}
\end{equation}
where $D_{ij}$ are the angular diameter distances between the observer
$o$, lens $l$  and source $s$. For solar  mass objects at cosmological
distances,   this  is   of   order  $\sim0.01$   parsecs,  and   hence
inhomogeneities  in the  absorbing  material would  have  to occur  on
scales substantially smaller than this to influence the individual light 
paths and hence modify the absorption line profile.

\begin{figure*}
\centerline{ \psfig{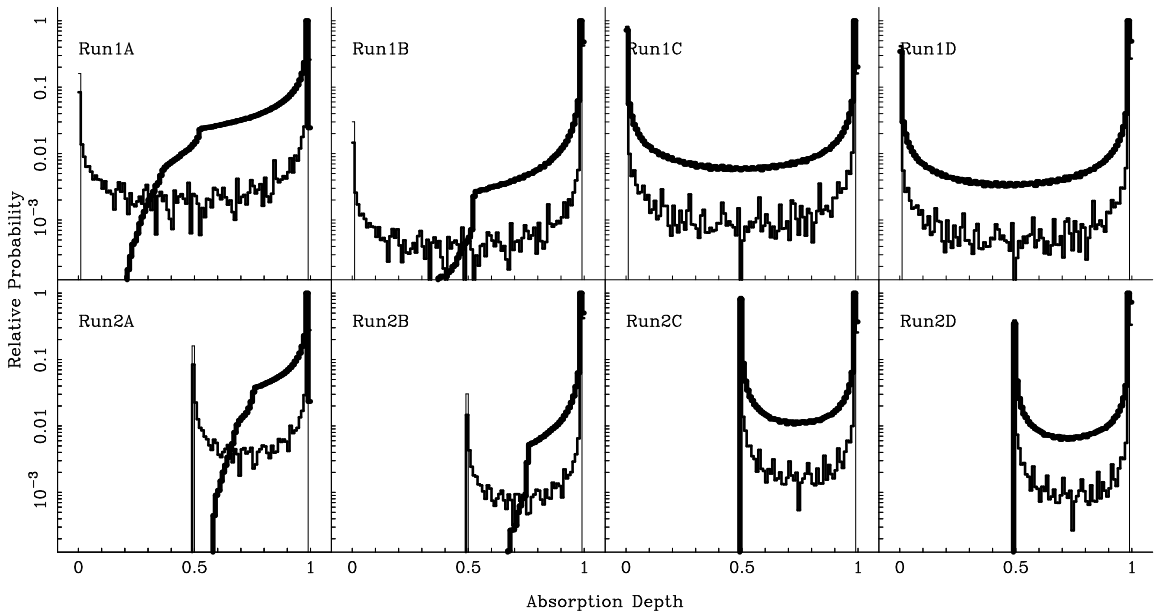}}
\caption[]{The distribution of  relative probabilities of a particular
absorption depth  occurring in  the for a  source observed  through the
cloud  distributions presented  in  Figure~\ref{clouds}.  While  three
sources  are considered in  the text,  only two  curves appear  on the
plots, indicating only the largest (1ER radii - thick line) and medium
(0.2ER  - light line)  sources appear  in the  plots, as  the smallest
source (0.02ER) produces a binary  distribution at the extremes of the
absorption range.
\label{Probs}}
\end{figure*}

\subsection{High optical depth microlensing}\label{high}
The light  from macro imaged  quasars shines through  relatively dense
regions of the lensing galaxy  where the optical depth to microlensing
is of order unity. In this regime, many stars influence the light from
the quasar  and the resulting lensing is  correspondingly more complex
than  that  seen for  an  isolated mass.   The  details  of these  are
described in  several recent reviews  and will not be  reproduced here
(e.g. Wambsganss 2001).

\begin{figure}
\centerline{ \psfig{figure=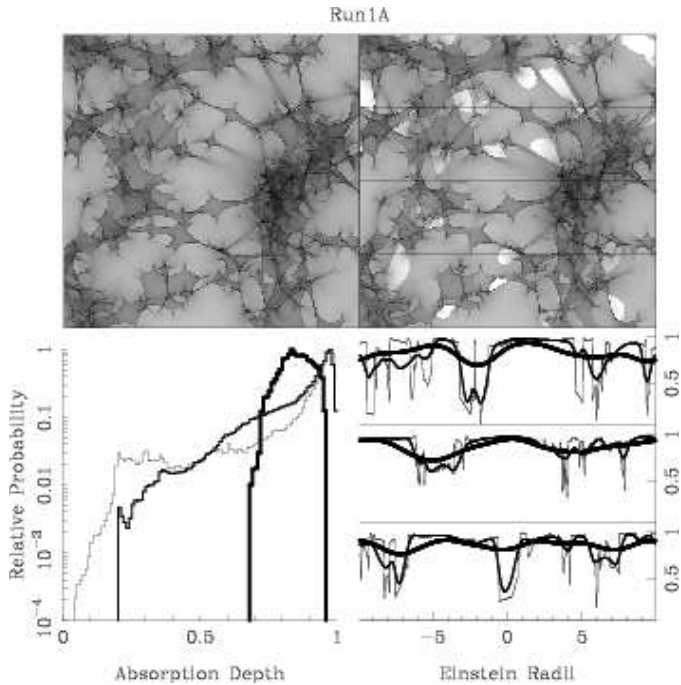,angle=270,width=3.5in}}
\caption[]{\label{run1a}  The  simulation   results  for  Run1A.   The
upper-left hand  panel presents the microlensing  magnification map in
the source  plane.  In the upper-right  hand map, the  presence of the
absorbing material is  considered and any ray passing  through a cloud
in  the  lens  plane  is  denuded  by  the  appropriate  amount.   The
lower-left hand panel presents the normalised probability distribution
of the  depth of the absorption  line (which is the  ratio between the
upper two magnification maps).  On the lower-right hand panel, several
light curves of the absorption depth are presented. Three source sizes
are  considered,  corresponding  to   the  three  lines  of  differing
thickness presented in the  lower panels. The thinest line corresponds
to the smallest  source, while the thickest is  the largest. Note that
the magnification  maps and  the duration of  the light curves  are 20
Einstein radii  (ER). In  this example, there  are 391 clouds  per 100
ER$^2$ with  a radius of 1ER.  The clouds are  opaque, transmitting no
radiation.}
\end{figure}

At  such high  optical  depths, the  observed  macroimage is  actually
composed  of  an  unresolved  myriad  of  microimages.   The  relative
brightnesses of  these images change as the  microlensing stars change
their  position in  front of  the  source.  Recently,  Lewis \&  Ibata
(1998) examined the astrometric shift in the macroimage as a result of
the changing  microimage configuration, finding  appreciable shifts of
milliarcseconds; such shifts will  be readily observable with the next
generation of space-based interferometers.

\begin{figure}
\centerline{ \psfig{figure=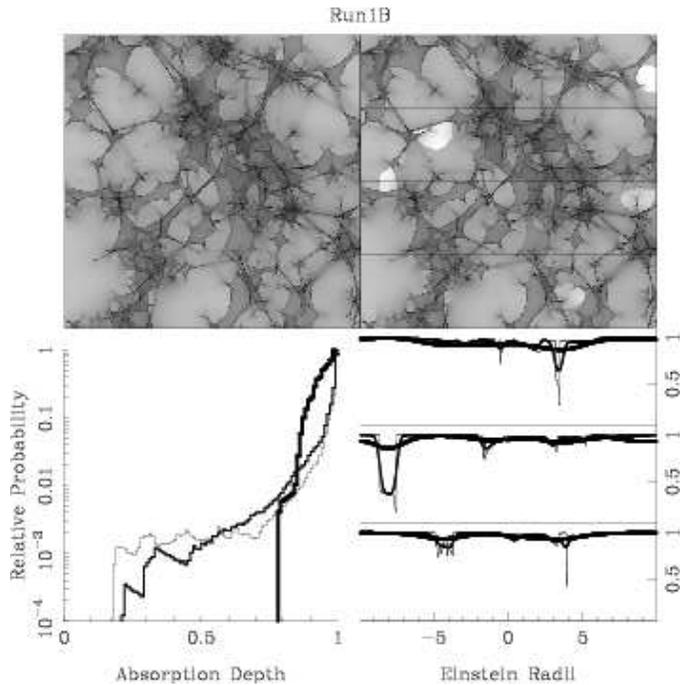,angle=270,width=3.5in}}
\caption[]{\label{run1b}As  for  Figure~\ref{run1a},  but  for  Run1B.
Here, there  are 40  clouds per  100ER$^2$ with a  radius of  1ER. The
clouds are opaque, transmitting no radiation.  }
\end{figure}

The  numerical approach  of Lewis  \& Ibata  (1998) was  to  image the
quasar source as it shines through an ensemble of microlensing masses;
from this  the brightness weighted  macroimage centroid can  be simply
calculated.   It is,  however,  quite straightforward  to extend  this
approach  to consider  the influence  of a  distribution  of observing
material  amongst the  lensing masses;  such an  approach  was adopted
here.

\begin{figure}
\centerline{ \psfig{figure=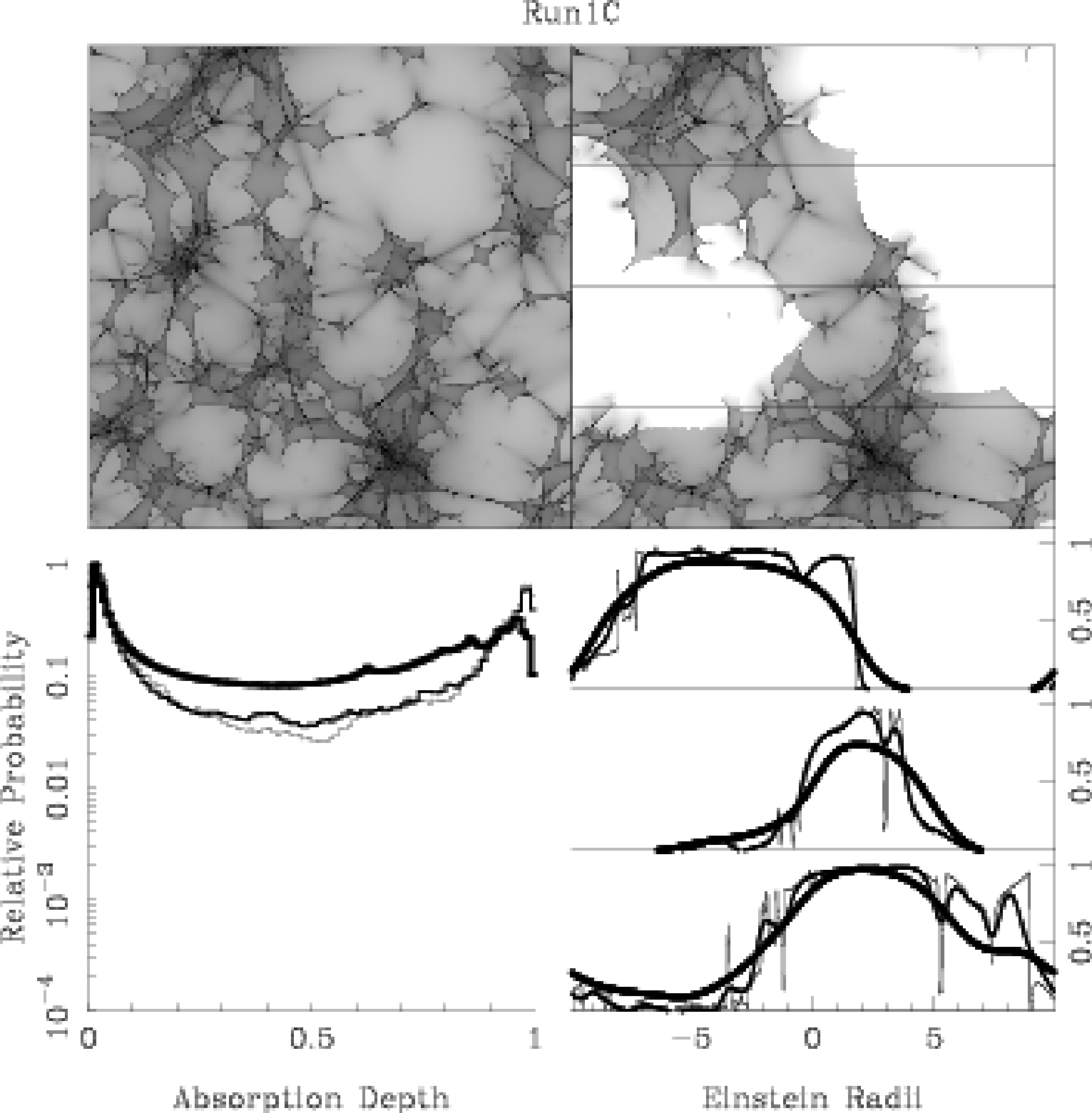,angle=270,width=3.5in}}
\caption[]{\label{run1c}As for Figure~\ref{run1a}, but for Run1C.
Here, there  are 31  clouds per  100ER$^2$ with a  radius of  8ER. The
clouds are opaque, transmitting no radiation. 
}
\end{figure}

In  summary,  the numerical  routine  was  based  on the  ray  tracing
procedure which has been the  work horse in the study of gravitational
microlensing (e.g. Kayser, Refsdal \&  Stabell 1986), in which a large
number  of rays  are fired  through  a population  of lensing  masses.
Considering where each ray intercepts the source plane, a microlensing
magnification map  can be derived  and the statistics  of microlensing
induced variability inferred (e.g.  Wambsganss 1992). For the purposes
of this  study, absorbing material  is added between  the microlensing
masses, attenuating the rays as  they pass through the lensing galaxy.
The  presence of  absorbing material  will influence  the form  of the
magnification map, as  regions will receive less rays  than they would
without it.  Hence, a  comparison between the `raw' magnification map,
with  no  absorbing  material,  and   that  in  the  presence  of  the
attenuating screen, gives the relative strength of the absorption line
to  the continuum.   It is  assumed  that the  clouds are  effectively
massless,   not   contributing   to   any   additional   gravitational
microlensing  convergence  (Wambsganss  1992;  Lewis  \&  Irwin  1995;
Schechter \& Wambsganss 2002).

\begin{figure}
\centerline{ \psfig{figure=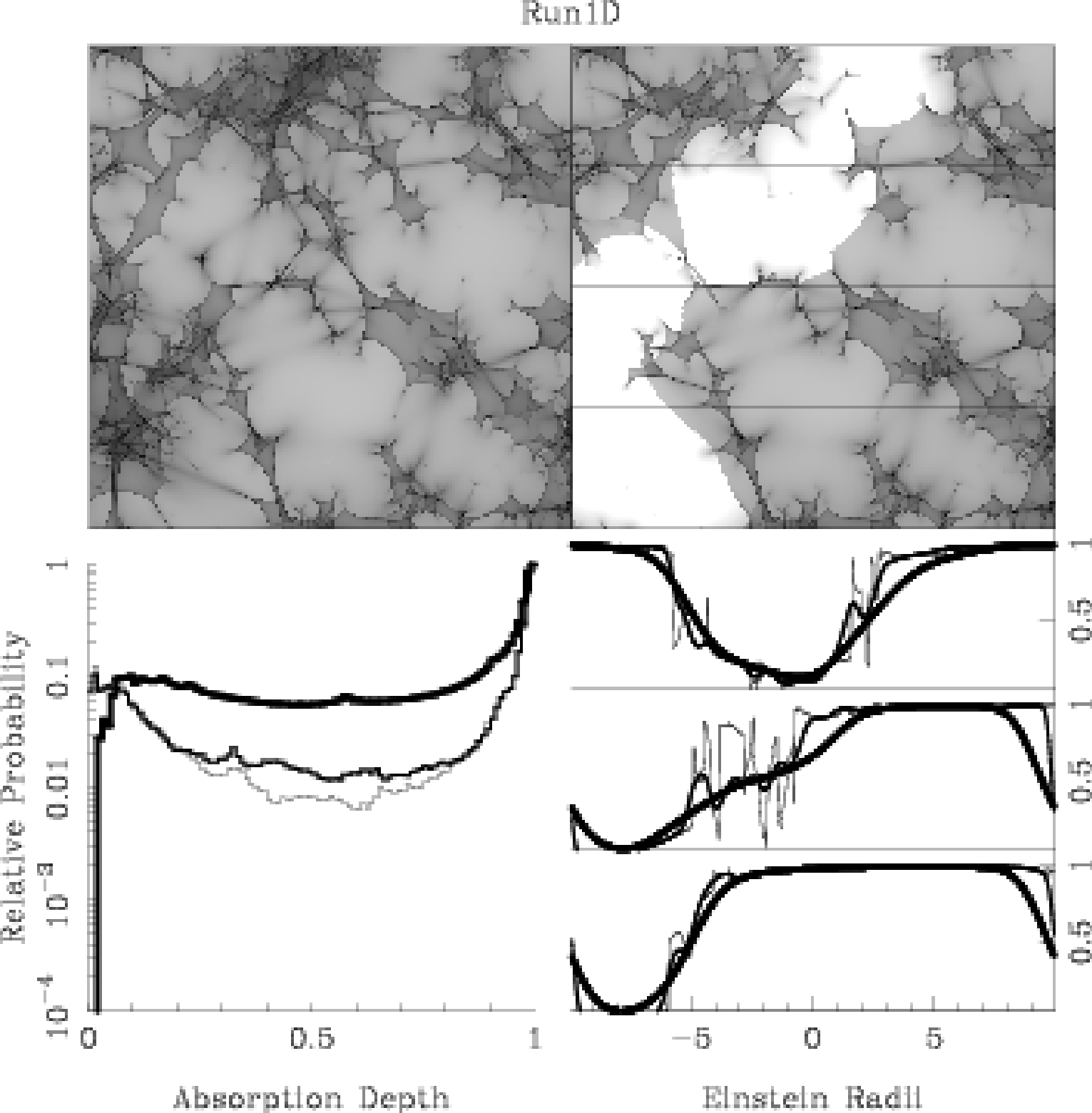,angle=270,width=3.5in}}
\caption[]{\label{run1d}As  for  Figure~\ref{run1a},  but  for  Run1D.
Here, there  are 15.5 clouds per  100ER$^2$ with a radius  of 8ER. The
clouds are opaque, transmitting no radiation. }
\end{figure}

The complex patterns  seen in gravitational microlensing magnification
maps depend  strongly on the  surface density of  microlensing objects
(the optical depth  $\sigma$) and the large scale  shear over the star
field (the  shear $\gamma$).  Furthermore, for  a particular macrolens
and  source, where  the physical  scales are  fixed, varying  the mass
function of the  microlensing masses also changes the  patterns of the
magnfication maps (Kayser, Refsdal  \& Stabell 1986; Wambsganss 1992).
Hence,  the   potential  parameter   space  available  for   study  is
considerable. For  this study,  a fiducial microlensing  parameters of
$\sigma=0.5$  and  $\gamma=0.0$ were  employed,  considering a  source
region which  is 20 Einstein  radii in extent. The  corresponding star
field is scattered over a  region of radius $\sim70$ Einstein radii in
the lens plane.

\begin{figure}
\centerline{ \psfig{figure=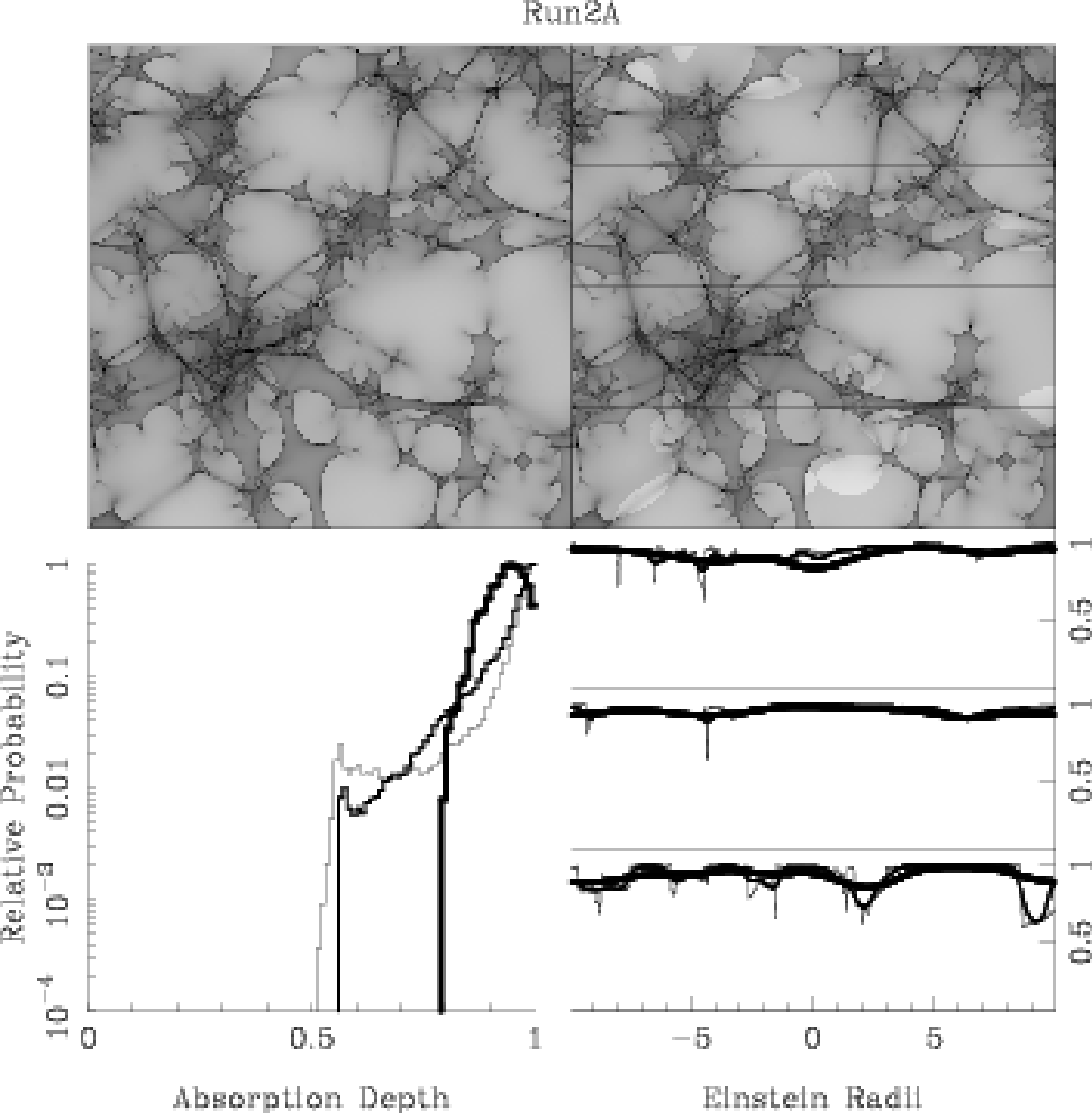,angle=270,width=3.5in}}
\caption[]{\label{run2a}As  for  Figure~\ref{run1a},  but  for  Run2A.
Here, there  are 391 clouds  per 100ER$^2$ with  a radius of  1ER. The
clouds transmit 50\% of radiation.  }
\end{figure}

\section{Absorption Cloud Distribution}\label{clouds}
Absorption  clouds can  be distributed  in a  multitude of  ways, with
varying shapes  and sizes and  absorption profiles etc. and  hence the
potential  parameter  space  to  be  explored  is  immense.   Such  an
exploration  is beyond  this  current contribution,  and  so, for  the
purposes of  this study, the absorption clouds  are simply represented
as being  circular with uniform absorption.  Two  fiducial cloud sizes
were considered,  the larger possessing  a radius of 8  Einstein radii
(ER), the smaller clouds  being $\sim1$ Einstein radii.  Two densities
of each  cloud distribution were  also considered.  Model  A possesses
small clouds  scattered with  a density of  391 clouds per  100 square
Einstein radii,  model B also has  small clouds, but  scattered with a
density of 40 clouds per  100 square Einstein radii.  Model C utilises
large clouds, with  a density of 31 clouds per  100 Einstein radii and
model D again  uses large clouds with a density half  that of model C.
Figure~\ref{fig3}  presents examples of  the four  cloud distributions
employed in this study;  the grey-scale represents the distribution of
the absorbing  material, whereas the  points are the positions  of the
stars.   Furthermore,   the  clouds  were  considered   to  be  either
completely  opaque,  absorbing all  photons  (absorption  model A)  or
producing 50\% absorption (absorption model B).

\begin{figure}
\centerline{ \psfig{figure=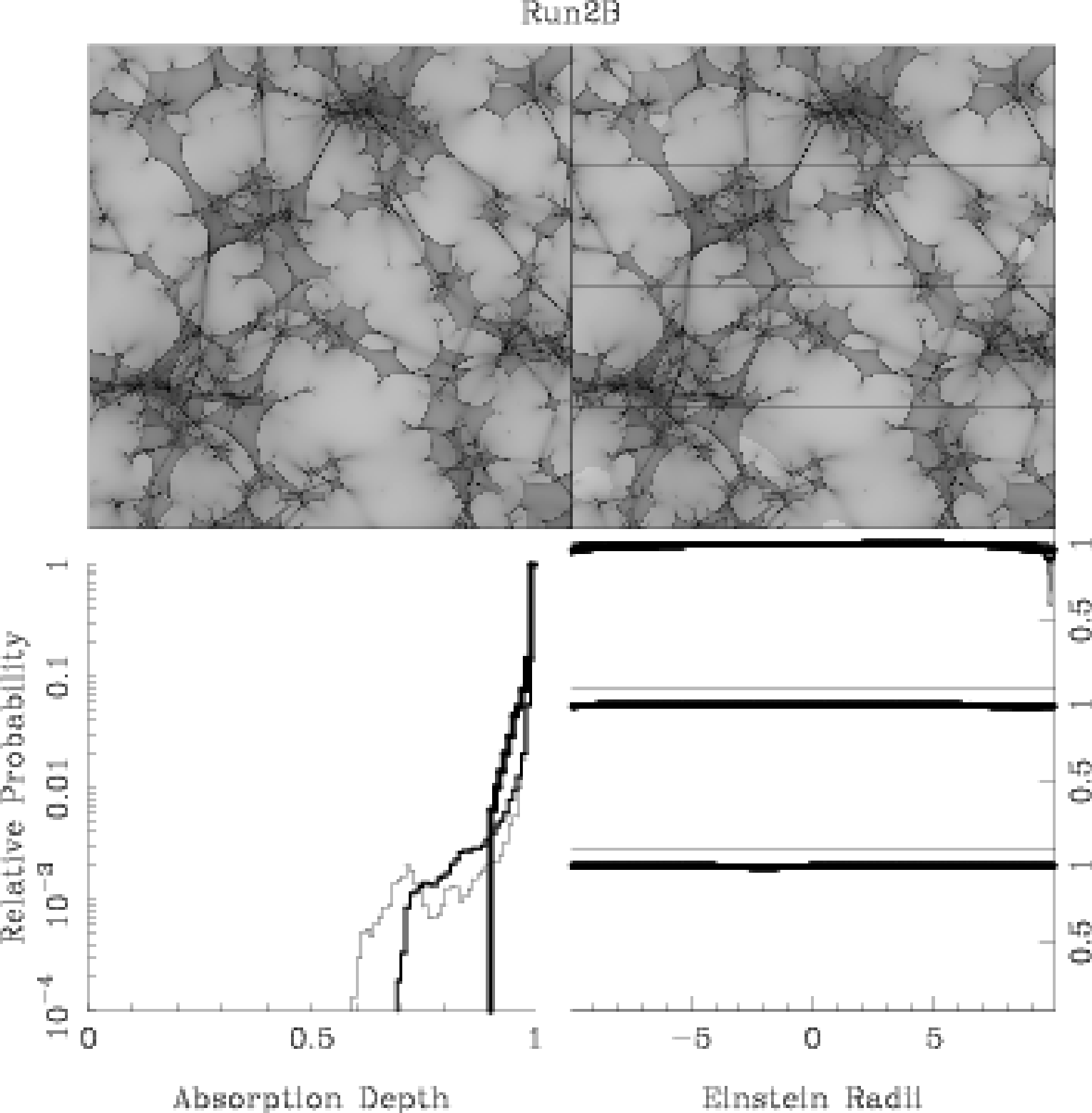,angle=270,width=3.5in}}
\caption[]{\label{run2b}As for Figure~\ref{run1a}, but for Run2B.
Here, there  are 40  clouds per  100ER$^2$ with a  radius of  1ER. The
clouds transmit 50\% of radiation.}
\end{figure}

Studies of the interstellar  medium reveal that clouds are distributed
fractally, possessing structure on  a range of scales (Elmegreen 1997).
When  sources possess  a fractal  distribution  it is  found that  the
scales of structure are imprinted on  the light curve as the source is
microlensed  (Lewis  2002). While  it  is  expected  that any  fractal
structure in the absorbing clouds will result in similar imprinting of
a scale of structure on the microlensing light curve, the calculations
required go beyond this study and are deferred for further work.

\begin{figure}
\centerline{ \psfig{figure=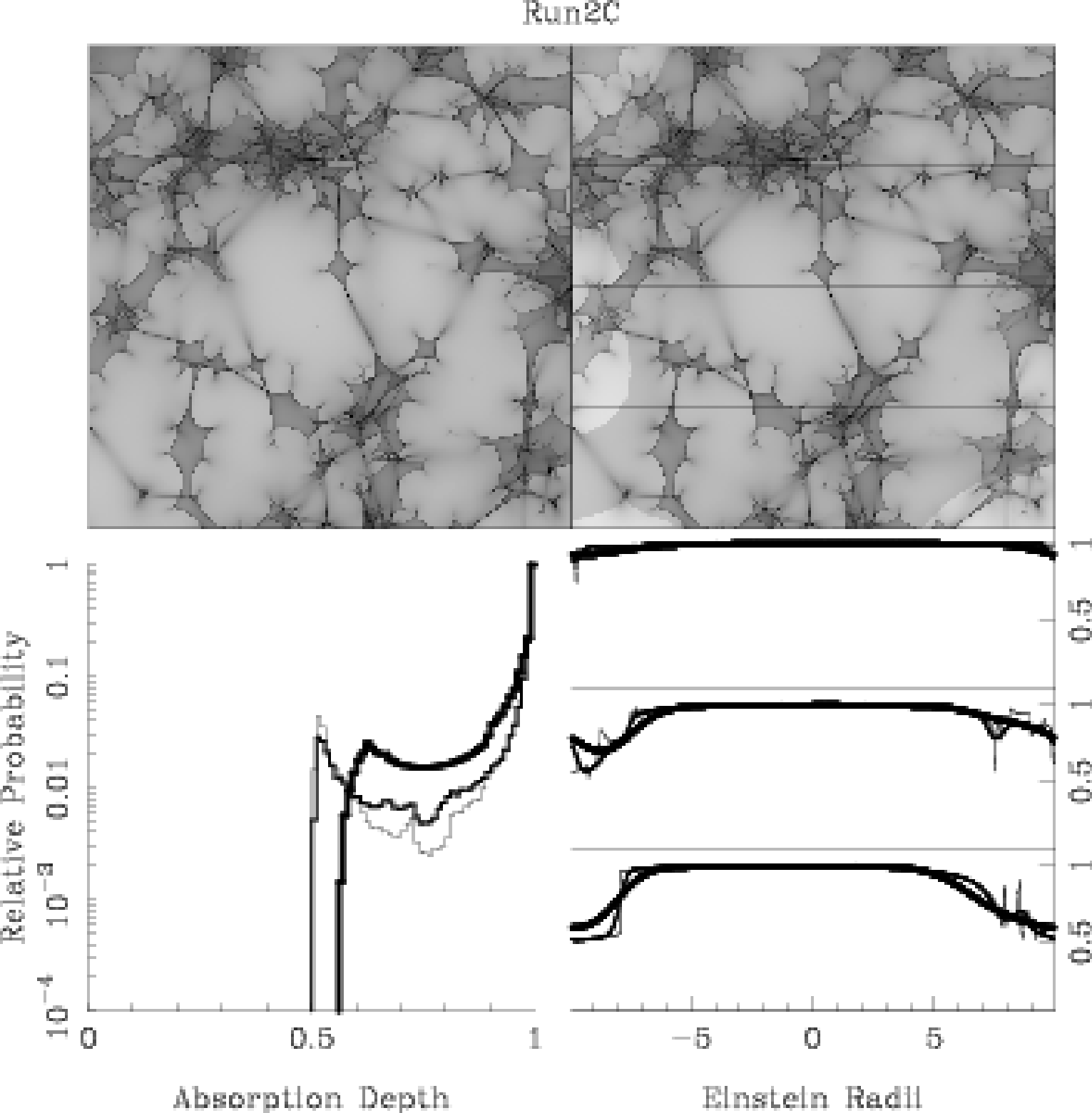,angle=270,width=3.5in}}
\caption[]{\label{run2c}As  for  Figure~\ref{run1a},  but  for  Run2C.
Here, there  are 31  clouds per  100ER$^2$ with a  radius of  8ER. The
clouds transmit 50\% of radiation.}
\end{figure}

\section{Magnification Maps}\label{magmaps}
\begin{figure}
\centerline{ \psfig{figure=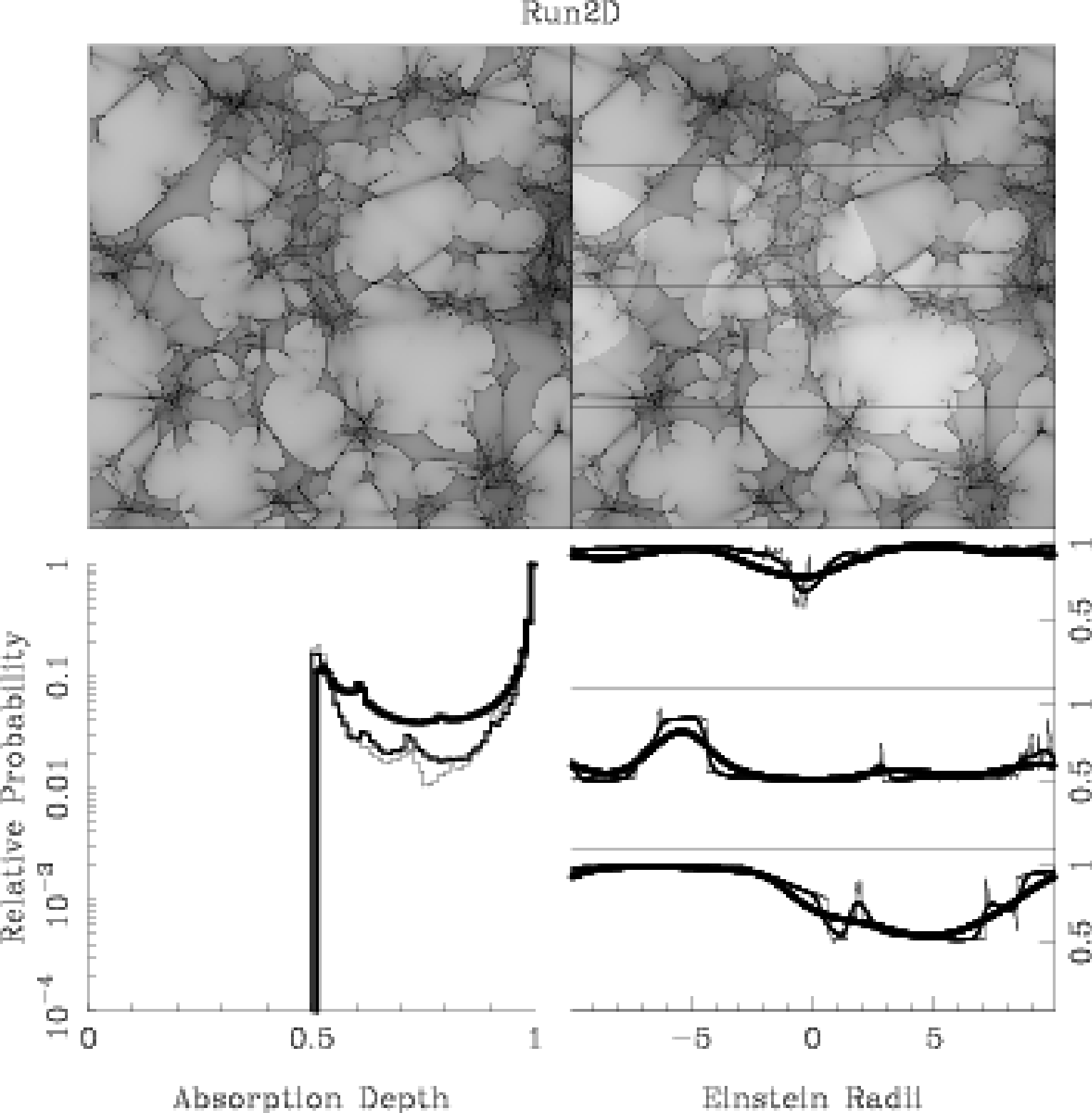,angle=270,width=3.5in}}
\caption[]{\label{run2d}As  for  Figure~\ref{run1a},  but  for  Run2D.
Here, there  are 15.5 clouds per  100ER$^2$ with a radius  of 8ER. The
clouds transmit 50\% of radiation.}
\end{figure}

Figures~\ref{run1a}  to   \ref{run2d}  present  the   results  of  the
numerical simulations detailed in the previous sections. The left hand
upper  panel  presents  the  microlensing  magnification  map  without
considering  the  influence of  absorbing  material,  with dark  areas
corresponding  to  regions of  magnification,  while  the light  areas
represent   regions   of   demagnification;   the   sharp   boundaries
corresponding  to  caustics  in  the  map are  clearly  visible.   The
upper-right hand panel again presents a magnification map, but in this
case any rays  which impinge on absorbing clouds  as they pass through
the  microlensing screen  are appropriately  denuded.  With  this, the
upper-left hand panel can be seen to be the magnification of continuum
emission,  while that in  the upper-right  hand panel  is that  in the
absorption line.  The relative depth  of the emission line seen at any
particular instant is the ratio of  these two maps. This can be simply
seen if one considers a  large screen of absorption that uniformly and
completely   covers   the   microlensing   star  field.   Hence,   the
magnification  map  in the  absorption  line  would uniformly  possess
values that are  a fixed fraction of those in  the continuum map, such
that the  ratio of the two  would be a constant  and would demonstrate
that,  in  this   case,  that  there  would  be   no  absorption  line
variability.

The  lower-right   hand  panel   presents  several  cuts   across  the
magnification  map  with  the   absorption.   There  are  three  paths
(indicated  by the  lines  across the  upper-right  hand panel).   The
absorption depth indicates how deep a line would cut into a normalised
continuum; a  value of one  indicated no absorption of  the continuum,
whereas 0  indicates complete extinction  in the line.  Three Gaussian
source sizes were considered: 0.02,  0.2 and 1 Einstein radii.  In the
figures, the lightest lines correspond to the smallest source, whereas
the darkest lines are the largest.

Finally, the lower-left hand panel  presents a relative measure of the
probability of  a particular  absorption occurring (normalised  to the
peak probability). Again, the  lightest line in this panel corresponds
to the smallest  source, and the darkest line is  that for the largest
source.

Neglecting the influence of  gravitational microlensing, what would be
expected if the  cloud distributions described in Section~\ref{clouds}
drifted across a source? If the source is smaller than the size of the
clouds,  then  either  source  will appear  unabsorbed  or  completely
covered  by an  absorbing  cloud.   If we  assume  that the  absorbing
material drifts  across the  sky with a  velocity V, then  the typical
time  scale  that  the  source  lies  behind  an  absorbing  cloud  is
$t_{abs}\sim R_{cloud}/V$. While completely behind a cloud, however, a
source  will  not display  any  absorption  line  variability, and  so
another important  time scale is  the time taken  for a cloud  to pass
from being unabsorbed  to completely absorbed; $t_{abs}\sim R_{src}/V$
for  a  source  smaller  than  the absorbing  cloud.   A  more  detail
discussion of the temporal properties of the light curve appears later
in this section.

The effects of  microlensing can be seen when  considering the top two
panels of each of the simulation outputs. When comparing the left-hand
panel (the  magnification map  without absorption) and  the right-hand
panel  (the same  map  but  including the  presence  of the  absorbing
material), it  is clear  that the presence  of the  absorbing material
denudes regions  of the magnification  map.  Due to  the gravitational
lensing  distortion, these  regions  are are  not  circular, like  the
absorbing clouds, but rather are distorted.  For the simulations where
large, opaque  clouds are considered (C  \& D), this  results in large
sections   of   the   magnification   map   being   essentially   zero
(Figures~\ref{run1c} \&  \ref{run1d}). When the opacity  of the clouds
is 50\%, the effect is less severe.

Examining Figures~\ref{run1a} to  \ref{run2d}, several features of the
absorption line variability are  apparent. Even with the limited range
of absorption cloud distributions presented in this paper, it is clear
that quite complex  variability in the strength of  an absorption line
results. As  with the strength of  microlensing induced magnification,
the degree  of variability is seen  to be strongly  dependent upon the
angular size  of the  source, with small  sources displaying  the most
most  dramatic  and   rapid  variability.   The  variability  strongly
reflects  the size  of the  clouds with  the larger  clouds  leaving a
longer imprint on the absorption  line light curve.  For virtually all
the scenarios  considered, the  source size was  substantially smaller
than the  absorbing cloud (the  exception being the largest  source in
models  A \& B,  where the  source is  comparable to  the size  of the
clouds).   Hence, if  the influence  of microlensing  is  neglected, a
source would  therefore find itself unobscured  by absorption material
or completely  behind a  cloud, with a  small transition  region. This
would lead to an  essentially binary distribution when considering the
absorption depth probabilities, with a  peak at one and another at the
absorption strength of the clouds (zero  for model 1 and 0.5 for model
2).  Figure~\ref{Probs}  presents   the  relative  probability  for  a
particular  absorption depth  occurring in  this  non-microlensed case,
clearly revealing the strong  binary nature of the expected absorption
depths.  As revealed in Figures~\ref{run1a} to \ref{run2d}, there is a
significant  probability in  all  cases to  get  absorption depths  in
between the  two extremes, with  quite different distributions  to the
non-microlensing case. In the non-microlensing case, absorption depths
at levels  between the  two potential extremes  can only occur  if the
source  lies  partly behind  an  absorbing  cloud. With  microlensing,
however,  this is  further augmented  by  the fact  that the  observed
macroimage is  composite and consists of many  microimages. Unless all
these  microimages suffer absorption  then the  observed depth  of the
absorption line  will be  less than a  single source viewed  through a
cloud.

In further addressing the influence of microlensing, it is instructive
to examine  the cloud distributions in  Figure~\ref{fig3}; the squares
in the centre of each plot  represent the source region over which the
magnification map  is calculated. Hence,  paths across this  square at
the same location as those in Figures~\ref{run2a} to \ref{run2d} (only
model 2 is considered as  they employ the specific cloud distributions
in  Figure~\ref{fig3}) represent  the  same light  curve durations  as
those presented in the microlensed  case. Firstly, it is apparent that
in  the absence  of microlensing,  Run2B and  Run2C should  exhibit no
absorption  line variability  as  they possess  no  clouds within  the
central  box. When  considering the  influence of  microlensing, Run2B
does  appear to show  little variability,  but significant  changes do
occur (as revealed by  the absorption depth probability distribution).
Run2C,   however,  shows  more   significant  variability,   with  the
gravitational microlensing effect `folding' the absorption information
into the  inner square.  An  examination of the cloud  distribution of
Run2D  suggests that, in  the absence  of microlensing,  the uppermost
path in Figure~\ref{run2d} should  be unobscured, while the lower path
should  be  completely  obscured.   This  is quite  different  to  the
behaviour  when  microlensing is  considered,  with  the light  curves
displaying more complex variability.

\begin{figure}
\centerline{ \psfig{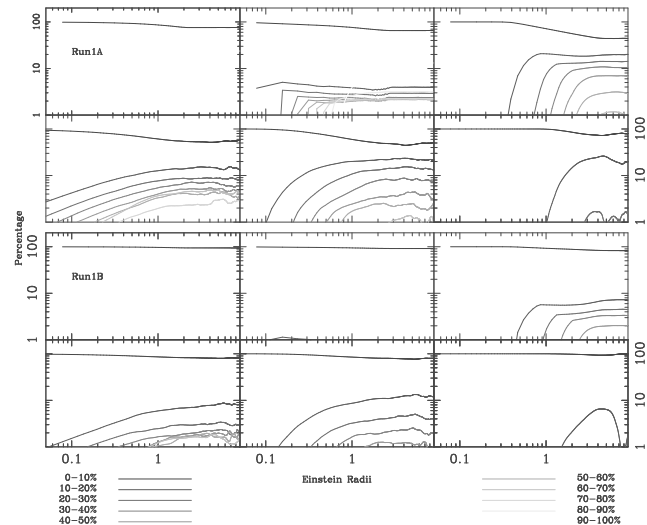}}
\caption[]{The temporal  properties of  the absorption line  depth for
the models Run1A and Run2B.  For each model, the results are presented
in two  sets of  three panels, the  upper panels for  an unmicrolensed
source  behind  the absorbing  clouds,  while  the  lower hand  panels
represent the case when the  source is microlensed. For each case, the
three panels,  from left to  right, represent the smallest  to largest
source.   The  curves  in  the  panels  represent  the  difference  in
absorption between two observations,  separated by a distance given on
the x-axis in  Einstein radii.  The y-axis presents  the percentage of
changes within a particular range, highlighted by the greyscale-coding
below the  plot. The  percentage is in  terms of the  maximal possible
change, being 1 for the first  four models and 0.5 for the second four
models.
\label{Temporal} }
\end{figure}

\begin{figure}
\centerline{ \psfig{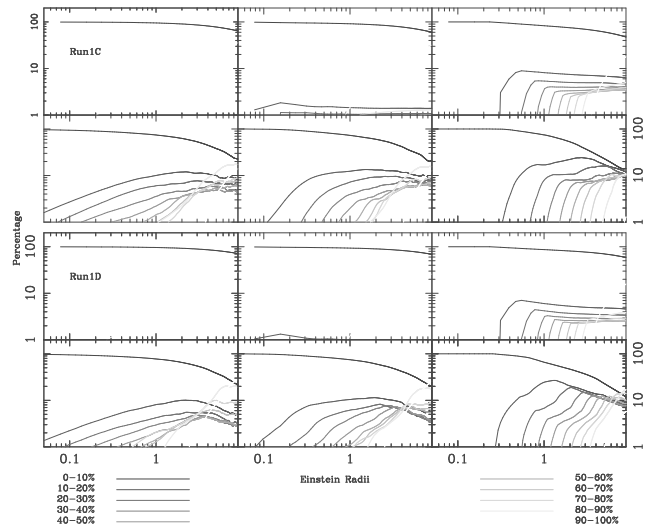}}
\caption{As Figure~\ref{Temporal}, but for models Run1C and Run1D.}
\label{Temporala}
\end{figure}

\begin{figure}
\centerline{ \psfig{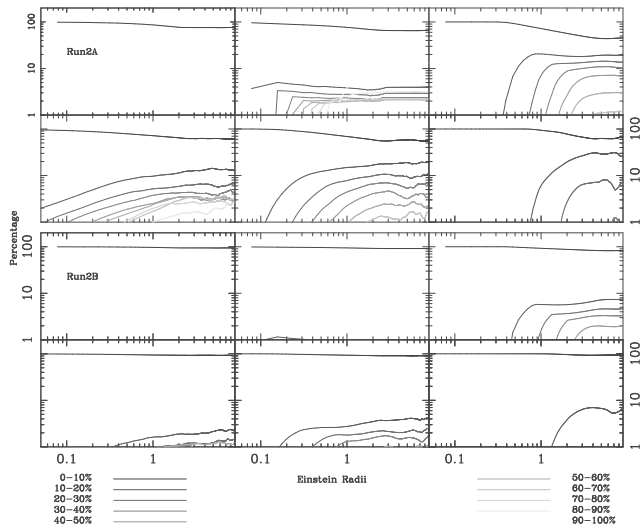}}
\caption{As Figure~\ref{Temporal}, but for models Run2A and Run2B.}
\label{Temporalb}
\end{figure}

\begin{figure}
\centerline{ \psfig{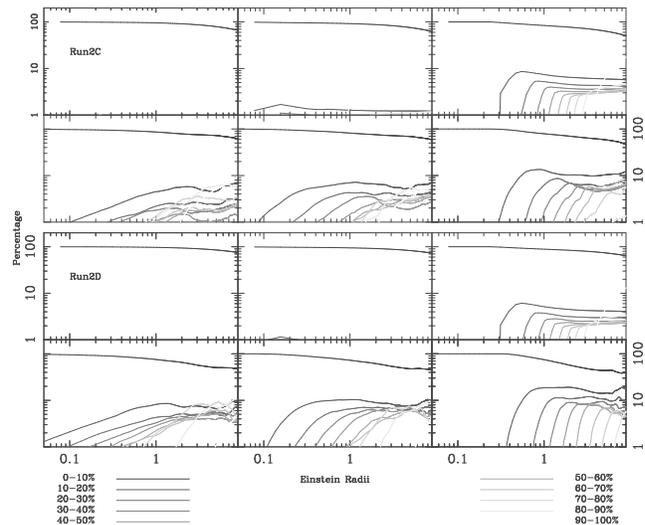}}
\caption{As Figure~\ref{Temporal}, but for models Run2C and Run2D.}
\label{Temporalc}
\end{figure}

In further examining the temporal  properties of light curves with and
without microlensing, a  study was made of the  expected change in the
absorption  depth as  a  function of  time;  as routine  spectroscopic
monitoring  of  lensed  systems   is  not  currently  undertaken,  any
observations are likely to be randomly spaced in time.  For this, this
study considers  the statistics of two measurements  of the absorption
line depth as  a function of their separation  in Einstein radii; this
separation  can  be converted  to  time  by  considering the  relative
velocities  of the source  and lens.   The results  of this  study are
presented  in  Figures~\ref{Temporal},\ref{Temporala}, \ref{Temporalb}
and \ref{Temporalc}; for each model (labeled at the far left) consists
of six panels.  The upper  three panels present the statistics for the
non-microlensing   cases,  the   lower  three   panels  are   for  the
microlensing case.  In each  set of three, the left-to-right subpanels
are  for the  smallest  to  largest source.   The  x-axis present  the
separation  in Einstein  radii; as  the  typical crossing  time of  an
Einstein radius of a cosmological  source is of order decades, a small
range  in Einstein  radius was  considered.  The  y-axis  presents the
percentage of  observations with this  separation displaying variation
within a particular range, denoted  by the greyscale key at the bottom
of  the plot;  note that  both axes  are logarithmic.   The percentage
changes are  in terms of the  maximal possible change,  which is unity
for models Run1, and 0.5 for models Run2.

Examining Figure~\ref{Temporal} reveals  that the temporal behaviour is
complex. Comparing firstly  the smallest sources in each  model, it is
clear that  observations at small separations  in the non-microlensing
case  are dominated  by the  smallest changes  in  absorption.  Within
increasing separation, however, the largest absorption change grows in
importance,  reflecting the  binary nature  of the  probability curves
presented  in Figure~\ref{Probs}.  Examining the  same panels  for the
microlensed  case we  see  that absorption  line  variability is  more
likely to be  seen at small separation (except  model Run2B), although
the differences are smaller than  the maximum; hence, for the smallest
source, more rapid variability will be apparent in the light curves of
microlensed  systems,  although at  a  value  less  than the  maximum.
Considering  the middle source  size, the  situation is  more complex.
Again,  the  shortest  time   scales  are  dominated  by  there  being
essentially no  change in the  observed absorption line depth  in both
cases.   For the  non-microlensed  case, however,  the several  models
display more  significant variability  into the shortest  time scales.
With  microlensing of  the middle  source  size, it  is apparent  that
significant  variation at the  very shortest  time scales  is unlikely,
although  the probability  of  observing more  significant changes  is
enhanced over the non-microlensing case. The situation for the largest
source size is  more complex, with all of  the non-microlensing models
displaying similar  shapes (although different  normalization) for the
various distribution.  With  microlensing, however, the models present
differing  distributions.  Examining  the  distributions presented  in
Figure~\ref{run1a}  through~\ref{run2d} it  is clear  that due  to the
very smooth nature of the light  curve for the largest source, with no
rapid  variability occurring  on short  time scales.   While,  from the
figure, longer time scale variability is apparent for such sources, on
the   shorter  time   scale  important   to   observational  programs,
significant variation will  not be seen for large  sources.  Hence, we
can conclude  that for the  smaller sources, microlensing leads  to an
enhancement in the possibility of observing a change in observations at
different epochs.

\begin{figure}
\centerline{ \psfig{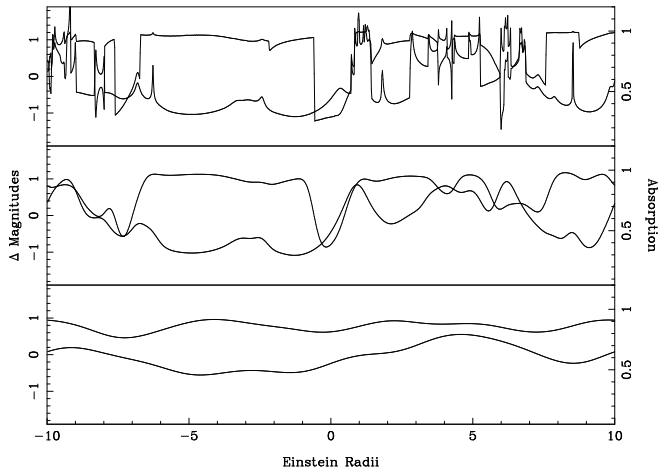}}
\caption[]{\label{SpecAbs} These  three panels present  the absorption
variation (lighter line) and light  curves (darker line) for the lower
path of Run1A (Fig.~\ref{run1a}). The smallest source is considered in
the upper panel, with the largest in the lower panel. It is clear that
while there  is some correlation  with the absorption  variability and
the light curve variations, this does not always hold.  }
\end{figure}

Is  there  a correlation  between  the  expected microlensing  induced
fluctuations  of the  continuum and  the expected  variability  in the
depth of  the absorption lines? When examining  the magnification maps
in  the  various  figures,  it  is  apparent that  the  value  of  the
magnification at  a particular  location is actually  the sum  of rays
that have  traversed differing  regions of the  lens plane.   With the
presence of absorbing material,  however, some rays being collected in
the  source plane  will be  attenuated.  A  closer examination  of the
magnification maps  reveals that some caustic structures  can lie over
regions  which have been  strongly absorbed.   Hence, we  would expect
some correlation  between the  continuum variability and  the observed
variations   in  the   absorption  depth.    This  can   be   seen  in
Figure~\ref{SpecAbs}  which presents the  light curves  and absorption
variability  for  the  lower  path  across the  magnification  map  in
Figure~\ref{run1a}.  The  top panel  presents the light  curve (darker
line)  and  absorption variability  (lighter  line)  for the  smallest
source considered, whereas  the lower panel presents the  same for the
largest source. It  is clear that, for the  smallest source, that some
changes in the observed absorption correlated strongly with changes in
the   continuum   brightness   (e.g.    rapid  change   in   both   at
$\sim$+1Einstein  radii), but  other  dramatic changes  in either  the
light  curve or  absorption  strength occur  unaccompanied. A  similar
trend is seen in the larger  source (middle panel), but in the largest
source considered, much  of the structure is washed  out and no strong
correlations occur.

This paper  has considered the simple situation  of comparing continuum
flux  with  that  at  the  centre  of an  absorption  line,  with  the
absorption line  profile simply scaling with this  central depth.  The
true  situation could  be more  complex, as  the absorbing  clouds are
likely  to   possess  a  range   of  velocities.   Hence,   the  cloud
distributions  presented in  Figure~\ref{clouds}  would represent  the
view at a specific wavelength, whereas other wavelengths may present a
different cloud  distribution. Hence, as well as  the overall strength
of the absorption line changing  due to the influence of gravitational
microlensing, the form of the line profile will also be time dependent
(e.g. Lewis \& Belle 1998).

\section{PKS 1830-211}\label{pks}
Jauncy  et al (1991)  resolved the  double radio  source \pks\  into a
ring-like  structure,   uncovering  its  gravitational   lens  nature.
Wilkind \&  Combes (1996) identified absorption due  to five different
molecular species, revealing  the lens to be at  a redshift of z=0.89,
whereas the  quasar source  was found  to be at  a redshift  of z=2.51
(Lidman et al.  1998).   Several approaches have successfully modelled
the gravitational  lensing potential (Kochanek \&  Narayan 1992; Nair,
Narashimha \& Rao 1993).  Recently, Winn et al.  (2002) and Courbin et
al.  (2002)~\footnote{It  should be noted  that these two  papers have
differing  interpretations  of the  same  data,  leading to  different
gravitational lens models.}  identified  the lensing galaxy in \pks, a
face-on spiral, in deep HST imaging; this naturally explains the large
quantities of molecular  gas seen in \pks\ as  the images are observed
through the spiral arms of the lensing galaxy.

As the  quasar images of \pks\ are  seen through the disk  of a spiral
galaxy suggests that our view  of this system may be further distorted
by the influence of  gravitational microlensing as stars stream across
the image of the quasar.   Lovell et al.  (1998) presented the results
of monitoring campaign  of \pks\ at 8.6GHz, from  which the time delay
between the  images was  determined to be  $\sim26$days. It  is clear,
however, that there are no  strong differences between the image light
curves that would suggest the influence of gravitational microlensing.
Oshima  et  al.   (2001),  however,  noted  that  the  relative  image
brightnesses in  \pks, as observed  with ASCA (Advanced  Satellite for
Cosmology  and   Astrophysics),  were  markedly   different  to  those
determined  from radio  observations,  leading them  to conclude  that
microlensing was  playing a  role in the  observed properties  of this
system. While  it appears that the  emission at 8.6GHz  comes from too
large  a region  to  be significantly  enhanced  by microlensing  (see
Wambsganss  1992;  however  Koopmans   \&  de  Bruyn  2000  identified
microlensing  at 8.5GHz in  CLASS B1600+434),  Garrett et  al.  (1997)
noted structural  changes, on  scales of $\sim1$milliarcsec,  in 15GHz
and  43GHz maps  taken  at  several epochs.   Garrett  et al.   (1997)
suggested that these variations were  due to millilensing by masses of
${\rm  \sim10^{4.5}M_\odot}$.   Interestingly,   Jin  et  al.   (1999)
observed \pks\ with the VLBA  at 43GHz over eight epochs, finding that
the  centroids of  the  core emission  varied significantly  ($\sim$80
microarcseconds over a period of  2 weeks). Similarly, a comparison to
archival  43GHz data  from one  year  earlier revealed  that the  core
separation had changed by $\sim$280microarcseconds.

Wiklind \&  Combes (1998) presented new observations  of the molecular
absorption lines  associated with the lens at  $z=0.89$.  These reveal
that the gravitationally lensed  images lie behind distinct absorption
clouds  separated  by   $\sim150$km/s.   When  comparing  to  previous
observations, Wiklind \& Combes  (1998) discovered that the profile of
the  ${\rm HCO^+(2 \leftarrow  1)}$ line  changed over  a $\sim6$month
time  scale;  while  these  changes  are relatively  small,  they  are
significant.

Is  the  change of  centroid  position  in  \pks\ consistent  with  it
undergoing  microlensing at  radio  wavelengths, and  if  so, is  this
responsible for  the observed absorption line  variability?  The Solar
mass   Einstein   radius   in   the   source  plane   for   \pks\   is
$\sim8\times10^{-3}h^{-\frac{1}{2}}$pc\footnote{The values have a weak
cosmological dependence.},  and sources must  be smaller than  this if
they are to be significantly  enhanced. This corresponds to an angular
scale of $\sim2$microarcseconds.  In their study of the quadruple lens
Q2237+0305, Lewis  \& Ibata  (1998) found that  image shifts  of 20-30
Einstein radii can occur on time scales substantially shorter than the
crossing time of  an Einstein radius. For \pks\  this crossing time is
$\sim33 h^{-\frac{1}{2}}$yrs, and the subsequent caustic crossing time
will be  of order  weeks to  months.  Hence, it  is expected  that the
separation  between the images  in \pks\  will change  by $\sim10-100$
microarcseconds on this time scale.  It should be noted, however, that
the  degree of  the  expected  astrometric shifts  are  a function  of
several  parameters,  especially the  macrolensing  shear.  Given  the
ring-like  nature  of  \pks\  (Jauncy   et  al.  1991),  this  may  be
substantial  in  the  vicinity  of  the images  of  the  quasar  core,
potentially accounting for the  large scale shifts observed by Garrett
et  al.   (1997).  More  detailed  simulations,  undertaken using  the
macrolensing  parameters for  \pks, are  required before  this  can be
fully addressed.

If  the  observed  line  profile   variability  in  \pks\  is  due  to
microlensing,   then   the   clouds   must  be   of   parsec/subparsec
scales. Rather  than reflecting individual clouds,  however, these may
represent inhomogeneities  in a larger  scale absorption distribution,
which could, potentially, be  fractal (Elmegreen 1997). It is apparent
that \pks\  is subject to  gravitational microlensing to  some degree,
although the  evidence is currently only suggestive  that the observed
line profile differences are  due to the absorption mechanism outlined
in this  paper.  Hence, further spectroscopic monitoring  of the radio
absorption  lines,  as  performed  by  Wiklind \&  Combes  (1998),  is
required.    Coupled  with   detailed   numerical  simulations,   such
monitoring may provide clues to the distribution of absorbing material
in galaxies on subparsec scales.

\section{Conclusions}\label{conclusions}
Employing  a  numerical  approach,  this paper  has  investigated  the
influence  of a  combination  of microlensing  and  a distribution  of
absorbing material within the  lensing galaxy upon temporal changes in
the depth of absorption lines in quasar spectra.  It is found that the
action  of macrolensing  compresses the  large scale  absorption cloud
information into  smaller region, as  seen by the source,  whereas the
microlensing   `folds'   the   absorption  patten.    Several   clouds
distributions were  examined, and significant modulation  of the depth
of  the absorption  line resulted.   The form  of the  absorption line
variability  differed significantly  from  the simple  case where  the
light from the quasar shone solely through an absorbing medium with no
additional   influence  from   gravitational  lensing.    Hence,  such
variability   could    influence   systems   suffering   gravitational
microlensing. Due to the small  size of the Einstein radius of stellar
mass objects at cosmological distances, however, only sub-parsec scale
variations in absorbing material will result in an observable effect.

The  gravitationally  lensed quasar,  \pks,  possess several  temporal
features  that   are  consistent  with   it  undergoing  gravitational
microlensing.   Furthermore,  \pks\ possesses  a  number of  prominent
molecular  absorption lines visible  at radio  wavelengths. In  one of
these,   ${\rm  HCO^+(2  \leftarrow   1)}$,  small,   but  significant
variations in the absorption line  profile over a period of six months
have been reported. The analysis presented in this paper suggests that
these  changes   may  be  due   to  the  influence   of  gravitational
microlensing.  It  is important to note, however,  that currently very
little is  known about  the scale of  structure of  absorbing material
within  external galaxies  and hence  the identification  of  any line
variability  as being due  to microlensing  may be  problematic. While
this paper has demonstrated  that microlensing induced absorption line
variability is  not perferctly correlated with the  variability of the
brightness  of the  macrolensed images,  some correlation  does exist.
Hence,  long  term photo-spectral  monitoring  at  high resolution  of
potential systems, such as \pks,  is required before such a conclusion
is confirmed.

\section*{Acknowledgements}
GFL  gratefully acknowledges discussions  with Joachim  Wambsganss and
Rachel Webster on aspects  of microlensing. Joachim Wambsganss is also
thanked for providing his {\tt  microlens} code on which this study is
based. We also thank Terry Bridges for allowing us to hammer odin, the
computer  on which  these  simulations were  undertaken. The  referee,
Stuart   Wyithe,  is   thanked   for  comment   which  improved   this
contribution.

\newcommand{\mnras}{MNRAS}
\newcommand{\nat}{Nature}
\newcommand{\araa}{ARAA}
\newcommand{\aj}{AJ}
\newcommand{\apj}{ApJ}
\newcommand{\apjl}{ApJ}
\newcommand{\apjs}{ApJSupp}
\newcommand{\aap}{A\&A}
\newcommand{\aaps}{A\&ASupp}

\end{document}